\documentclass[reprint,
prb,
floatfix,
aps,
nofootinbib,
]{revtex4-2}

\usepackage{graphicx}
\usepackage{multirow}
\usepackage{hhline}

\usepackage{physics}
\usepackage{siunitx}

\usepackage[colorlinks=false]{hyperref}
\usepackage{xcolor}

\begin{document}
\author{Jason Marfey}
\email{jmarfey3@gatech.edu}
\author{Anthony Semenova}
\email{ksa6@gatech.edu}
\author{Colin V. Parker}
\email{cparker@gatech.edu}

\affiliation{School of Physics, Georgia Institute of Technology, Atlanta, GA, 30332, USA}

\title{Crystal Fields and Zeeman Effect for Thulium in Solid Argon}

\begin{abstract}
    Optically active defects suspended in an inert solid are an interesting system for sensing and magnetometry at the nanometer scale,
    in addition to being a potential source of high-density, identical quantum emitters for quantum information.
    Beyond their response to external fields, the optical absorption and emission spectra also reflect information about the host matrix,
    which is critical to understand for either application.
    For the particular system of thulium atoms implanted in solid argon, high resolution laser spectroscopy reveals narrow ensemble linewidths of the \SI{1140}{\nano\meter} $f$ to $f$ ground state spin-orbit transition, which is split due to crystal field effects and hyperfine coupling.
    Pump-probe spectroscopy is used to identify crystal field levels in at least two stable trapping sites, and the crystal field is determined to be nearly axial in both sites.
    Strong selection rules indicate that this transition becomes magnetic-electric in the argon host, most likely due to dielectric effects.
    In the presence of \SI{}{\milli\tesla} magnetic fields, Zeeman shifts are resolvable by laser fluorescence, allowing crystal axis selective polarized spectroscopy.
    These results show that all-optical detection schemes for DC magnetic fields are already possible with thulium-doped argon,
    and point towards more informed strategies for monitoring and improving crystal growth and annealing.
\end{abstract}

\maketitle

\onecolumngrid
\begin{center}
Published in Phys. Rev. B \textbf{112}, 014101 (2025).\\ \copyright 2025 American Physical Society. \href{https://doi.org/10.1103/x5zx-ykff}{https://doi.org/10.1103/x5zx-ykff}
\end{center}
\vspace{1em}
\twocolumngrid

\section{\label{sec:introduction}Introduction}

Solids formed by the rare gases at cryogenic temperatures have a number of qualities making them an ideal host for optically active defects:
they are chemically inert, transparent into high UV, and retain the guest species for long periods of time.
Historically, these properties have been useful for spectroscopic study of reactive species~\cite{Becker_1956} and the rovibrational states of molecules~\cite{Fajardo_2009}.
Recently, systems isolated in a rare-gas matrix have been investigated for tests of fundamental physics~\cite{Braggio_2022, Arndt_1993, Vutha_2018} and quantum sensing~\cite{Lancaster_2024}.
In particular, matrix-isolated atoms have advantages as nanoscale sensors because they possess no charge and can be grown in the inert solid by vapor deposition, allowing co-trapping of target and sensor species or growth near a surface.

For any of these applications, precision matrix isolation spectroscopy experiments are necessary to understand the effects of the rare-gas host matrix on the atom, and many recent studies have been devoted to the alkali~\cite{Fajardo_1991, Tam_1993, Battard_2023, Lancaster_2021} and transition~\cite{Collier_2010, Bracken_1997} metals.
Effects of the matrix on the defect are relatively weak, originating from Van der Waals forces, but nonetheless shift and broaden atomic spectral lines at the scale of \SI{}{\tera\hertz}.
Atomic spectra are further complicated by crystal fields, Jahn-Teller effects, and multiple trapping sites, combinations of which must be used to understand absorption and emission spectra.
Weak matrix effects make the spectroscopy of matrix-isolated lanthanides interesting due to the shielding of the unfilled $f$ shell by the filled $5s$, $5p$, and $6s$ shells.
Of the lanthanides studied~\cite{Xu_2011, Byrne_2011_2}, the thulium atom ($Z = 69$) is host to a ground state transition at $\SI{1140}{\nano\meter}$ which occurs between spin-orbit levels contained within the $\text{[Xe]}4f^{13}6s^2$ ground configuration.
This ``inner-shell'' transition is naturally resilient to the environment~\cite{Golovizin_2019}, and has been shown to remain narrow and unshifted in liquid helium~\cite{Ishikawa_1997} and in solid argon and neon~\cite{Gaire_2019}.
Direct laser fluorescence spectroscopy of this transition has shown that it is split into multiple components with narrow ensemble linewidths and low inhomogeneous broadening~\cite{Gaire_2023}.

In this work, we investigate this ${}^2F_{7/2}$ to ${}^2F_{5/2}$ ground state transition of thulium in solid argon, identifying at least two thermally stable trapping sites using pump-probe fluorescence spectroscopy.
In both sites, the degeneracy is lifted by the crystal field, and the measured energy structure is determined mostly from a purely axial crystal field, allowing line assignments to be made.
The selection rules across both sites indicate that the gas phase magnetic dipole transition becomes mixed magnetic-electric in the argon matrix, most likely due to dielectric effects in the host.
With sample annealing, we observe extremely narrow, homogeneously broadened ensemble linewidths as low as $\SI{129}{\mega\hertz}$, enabling the direct resolution of both hyperfine structure and Zeeman shifts.
The hyperfine structure, which can be calculated from first principles, is used to confirm line assignments in one of the two stable trapping sites.
Zeeman shifts dependent on the relative orientation of the crystal axis with the applied field are used to coarsely resolve the crystal axis orientation, enabling polarized spectroscopy to confirm the transition type as magnetic-electric.

Narrow ensemble hyperfine lines sensitive to Zeeman splitting enable an all-optical magnetic field detection scheme, similar to those used in atomic vapor cells, on a smaller spatial scale allowed by the solid-state host.
Nanoscale magnetometry is typically performed using the spin-dependent photoluminescence of nitrogen vacancy (NV) centers in diamond~\cite{Rondin_2014}, but matrix-isolation magnetometry may ameliorate the limitations of sensor proximity, charge instability, and surface field noise suffered by NV magnetometers~\cite{Schirhagl_2014}.
Optically detected magnetic resonance with long coherence times has been observed for alkalis in inert matrices~\cite{Kinoshita_1994, Moroshkin_2006,Upadhyay_2019,Lancaster_2024}, using detection protols similar to those in NV centers, relying on optical pumping and resonant radio frequencies to excite the ground state spin configuration.
All-optical techniques, which eliminate the requirement to bring radio frequency radiation onto the sample, have also been developed for NV centers~\cite{Burgler_2023, Horsthemke_2024, Rondin_2012,Wickenbrock_2016}, but have not existed in rare-gas hosts.
The narrow ${}^2F_{7/2}$ to ${}^2F_{5/2}$ transition in thulium-doped solid argon introduces a new paradigm for all-optical magnetic sensing with matrix-isolated species, allowing detection of \SI{}{\milli\tesla} fields directly from the optical spectra.

\section{Experiment}

The procedure for the growth of thulium-doped rare-gas solids has been described previously~\cite{Gaire_2019, Gaire_2023}.
Briefly, a copper sample holder containing the tip of an optical fiber is mounted to the cold head of a closed-cycle cryostat capable of reaching \SI{4}{\kelvin} (with the fiber tip likely a few \SI{}{\kelvin} hotter).
Argon gas is directed through a 1/16" OD tube on the vacuum shroud and condenses on the fiber tip while the sample holder is held at \SI{15}{\kelvin}, with the growth rate measured to be \SI{135}{\nano\meter\per\second} by reflectometry of a \SI{662}{\nano\meter} laser diode.
Throughout the total growth time of 20 minutes, thulium is supplied by focusing a pulsed \SI{532}{\nano\meter} Nd:YAG laser beam onto a thulium metal sample which is rotated periodically.
After growth, the sample is annealed by resistive heating to a maximum of \SI{50}{\kelvin} for short durations, during which there is considerable risk of sublimation or sample detachment.
A single sampled held at base temperature remains stable for a duration of months and is removed only by deliberate heating.

\section{\label{sec:fluorescence}Fluorescence Spectroscopy}

\begin{figure*}
    \centering
    \includegraphics[width=\linewidth]{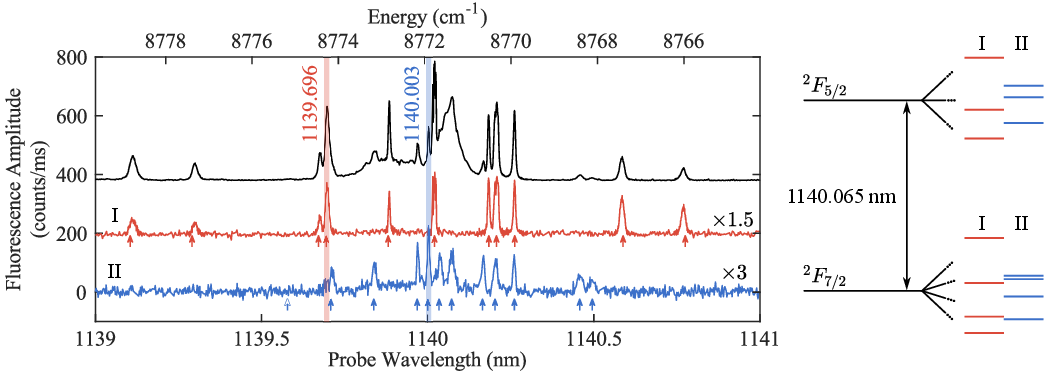}
    \caption{\label{fig:pumpprobe} Infrared fluorescence spectrum of the ${}^2F_{5/2}\leftrightarrow{}^2F_{7/2}$ transition of Tm in Ar. The black curve is a typical fluorescence spectrum obtained for a sample annealed to \SI{35}{\kelvin}. (I, red underset curve) Reduction in probe fluorescence with additional pumping at \SI{1139.696}{\nano\meter}. (II, blue underset curves) Reduction in probe fluorescence with additional pumping at \SI{1140.003}{\nano\meter}. Arrows indicate feature assignments corresponding to transitions between the energy levels displayed at right for, I, \SI{1139.696}{\nano\meter}, and II, \SI{1140.003}{\nano\meter} pumping. The empty arrow indicates a forbidden transition for II (the corresponding forbidden transition is outside figure bounds for I).}
\end{figure*}

A tunable external cavity diode laser scanned from $\qtyrange{1139}{1141}{\nano\meter}$ was used to obtain the fluorescence spectrum of the ${}^2F_{7/2} \leftrightarrow {}^2F_{5/2}$ thulium ground state transition in solid argon.
A resonant \SI{4}{\milli\second}, \SI{20}{\milli\watt} pulse was delivered through optical fiber, and the resulting fluorescence was measured for \SI{100}{\milli\second} through the same fiber.
The fluorescence profile was fit to a single exponential decay at each wavelength, the amplitudes of which are shown in Figure~\ref{fig:pumpprobe} for a sample annealed to \SI{35}{\kelvin}.
The fluorescence spectrum contains at least 18 sharp features, whereas a fully lifted degeneracy of the $J = 7/2$ and $J = 5/2$ into Kramer's doublets could maximally produce 12 lines.
Thus, we investigate the presence of multiple trapping sites using two lasers in a pump-probe scheme, utilizing saturation effects which are shared across all related lines.
A strong \SI{40}{\milli\watt}, \SI{4}{\milli\second} pump pulse fixed on a sharp feature brings the targeted site close to saturation, after which an additional \SI{20}{\milli\watt}, \SI{4}{\milli\second} probe pulse produces a linear fluorescence response on unrelated lines and reduced fluorescence on saturated lines.
The reduction of fluorescence response due to the probe pulse (compared to the sum of separate pump and probe signals) is shown in Figure~\ref{fig:pumpprobe} for two pump wavelengths, \SI{1139.696}{\nano\meter} and \SI{1140.003}{\nano\meter}, belonging to ``Site I" and ``Site II", respectively.
The presence of at least two trapping sites is consistent with many matrix isolation experiments~\cite{Kleshchina_2019,Byrne_2011_2,Battard_2023} and predictions by theoretical studies~\cite{Ozerov_2021}.

Each site produces 11 lines, corresponding to full lifting of the $J = 7/2$ and $J = 5/2$ degeneracy with one forbidden transition (identified below as corresponding to $\abs{\Delta m_J} = 3$).
A combinatoric search for the best fit spacings of the sublevels based on the observed line positions yields those shown in Figure~\ref{fig:pumpprobe} and listed in Table~\ref{tab:ELevels}.
The transitions are centered at \SI{1140.065}{\nano\meter} for both sites, remarkably close to the \SI{1140.0896}{\nano\meter} gas-phase transition for neutral Tm~\cite{Golovizin_2019}, so we classify each site as containing a single neutral thulium atom.
Ablation byproducts like clusters or ions~\cite{Heimbrook_1987} and other trapping sites may be present, but do not contribute to the narrow and intense fluorescence in the scanned range.
The sites are further differentiated by radiative lifetime, \SI{30}{\milli\second} for Site I and \SI{24}{\milli\second} for Site II, and by visible emission profiles.
For Site I, multiple visible emission bands are present, presumably coming from two-photon and higher processes~\cite{Gaire_2019}, while no visible emission is observed for Site II.

\begin{table}[h]
    \centering
    \begin{tabular}{c|cccc}
        Site & Term & Energy (\SI{}{\per\centi\meter}) & Predicted (\SI{}{\per\centi\meter}) & $m_J$ \\
        \hline
         \multirow{7}{*}{I} & \multirow{4}{*}{${}^2F_{7/2}$}& $-3.530(3)$  & -3.367 & $\pm 1/2$\\
                        && $-2.089(3)$ & -2.020 & $\pm 3/2$\\
                        && $0.840(3)$ & 0.673 & $\pm 5/2$\\
                        && $4.780(8)$ & 4.714 & $\pm 7/2$\\
                        \\
          & \multirow{3}{*}{${}^2F_{5/2}$}& $8768.242(3)$ & 8768.195 & $\pm 1/2$\\
                        && $8770.744(5)$ & 8770.619 & $\pm 3/2$\\
                        && $8775.293(1)$ & 8775.467 & $\pm 5/2$\\
        \hline
         \multirow{7}{*}{II} & \multirow{4}{*}{${}^2F_{7/2}$}& $-2.31(2)$ & -2.0812 & $\pm 7/2$\\
                        && $-0.33(3)$ & -0.2973 & $\pm 5/2$\\
                        && $1.18(3)$ & 0.8920 & $\pm 3/2$\\
                        && $1.46(9)$ & 1.4866 & $\pm 1/2$\\
                        \\
            & \multirow{3}{*}{${}^2F_{5/2}$}& $8769.60(9)$ & 8769.6425 & $\pm 5/2$\\
                        && $8771.83(5)$ & 8771.7832 & $\pm 3/2$\\
                        && $8772.83(4)$ & 8772.8536 & $\pm 1/2$\\
    \end{tabular}
    \caption{
    Observed ${}^2F_{7/2}$ and ${}^2F_{5/2}$ crystal field levels obtained by a least squares combinatoric fit to the line positions shown in Figure~\ref{fig:pumpprobe}.
    Statistical uncertainties are determined from levels fit from: hyperfine structure (Site I), single Lorentzian (Site II), limited by wavemeter accuracy.
    Predicted crystal field energies are determined from an axial crystal field with parameters $\beta_2^0 = 0.3142$ (Site I),  $\beta_2^0 = -0.1433$ (Site II), with angular momentum assignments.
    }
    \label{tab:ELevels}
\end{table}

The crystal field reflects the point group symmetry of the local trapping environment, lifts degeneracies of the trapped atom, and can be represented with Stevens operator equivalents whose coefficients are the crystal field parameters.
We modeled the crystal field with the Stevens operator $O_2^0 = 3L_z^2 - L^2$, defined with $L = 3$ for $f$ electrons whose spin does not couple to the crystal symmetry~\cite{Stevens_1952}.
Choosing the $O_2^0$ Stevens operator reproduces the relative scaling of $9/5$ between the $J = 7/2$ and $J = 5/2$ manifolds and the interval rule which is roughly obeyed for each site, as shown in Figure~\ref{fig:pumpprobe} and Table~\ref{tab:ELevels}.
A single parameter fit to the $O_2^0$ perturbation produces the crystal field parameters $\beta_2^0 = 0.3142$ for Site I and $\beta_2^0 = -0.1433$ for Site II, and the predicted level structure is shown alongside observed values in Table~\ref{tab:ELevels}.
The $O_2^0$ operator preserves $m_J$ as a good quantum number, giving line assignments shown in Table~\ref{tab:ELevels}.

Inclusion of higher order crystal field terms only marginally improves the agreement of the predicted levels, and the low order, axially symmetric Stevens operator plays a dominant role, implying a trapping site symmetry lower than tetrahedral.
Spectroscopic and theoretical studies on the similar Yb ($Z=70$) in argon identify single-substitution, tetravacancy, and hexavacancy trapping sites, corresponding to cubo-ocothedral, tetrahedral, and octahedral symmetries~\cite{Tao_2015,Ozerov_2019}; similar trapping sites have also been proposed for matrix isolated Europium ($Z = 63$)~\cite{Byrne_2011_1, Byrne_2011_2}.
For thulium atoms, the trapping site symmetry must be lower such that the $J = 7/2$ and $J = 5/2$ degeneracies are lifted.

Generally, the multiplet structure of matrix-isolated atoms also includes Jahn-Teller effects~\cite{Crepin-Gilbert_1999}, both static and dynamic, which spontaneously break the degeneracy of symmetric electronic levels and induce a lattice distortion.
However, the observed splitting is on the order of \SI{}{\per\centi\meter}, while the Debye energy for other atoms isolated in argon is typically around \SI{30}{\per\centi\meter}~\cite{Collier_2010, Healy_1999, Kleshchina_2019, Battard_2023}, nearly an order of magnitude larger.
Combined with the lack of clear phonon lines, this supports the suitability of a purely crystal field model.

\section{Hyperfine Structure And Selection Rules\label{sec:hyperfine}} 

\begin{figure}
    \centering
    \includegraphics[width=\linewidth]{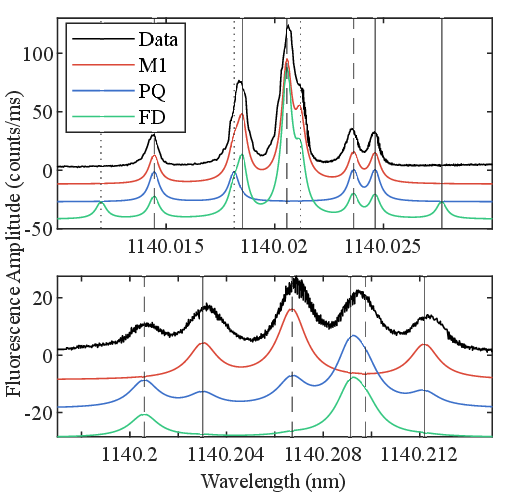}
    \caption{Measured hyperfine spectra along with predicted hyperfine amplitudes for M1 (magnetic dipole), PQ (pseudoquadrupole), and FD (forced dipole) transitions. Top and bottom are the $\abs{m_J}: 1/2 \leftrightarrow 1/2$ and $\abs{m_J}: 1/2 \leftrightarrow 3/2$ transitions, respectively. Styled vertical lines indicate hyperfine splittings determined from Reference~\cite{Mishin_2024}, with lines of the same style originating in the same ${}^2F_{7/2}$ ground state hyperfine level.}
    \label{fig:hyperfine}
\end{figure}

Brief anneals of the Ar matrix to temperatures \SI{>45}{\kelvin} produce
absorption lines as narrow as \SI{0.56}{\pico\meter} (\SI{129}{\mega\hertz}), allowing much of the hyperfine structure to be resolved for Site I.
Figure~\ref{fig:hyperfine} shows the optically resolved hyperfine structure of the $\abs{m_J} : 1/2\leftrightarrow1/2$ and $\abs{m_J} : 1/2 \leftrightarrow 3/2$ transitions along with predicted hyperfine splittings based on a pure axial crystal field and the hyperfine constants measured for Tm atoms in an optical lattice~\cite{Mishin_2024}.
Note that for given $\abs{m_J}$ this model is completely \textit{ab initio} with no fit parameters.
The excellent agreement confirms the assignments made from the crystal field in Section~\ref{sec:fluorescence}.

The presence of $\abs{\Delta m_J} = 2$ transitions (see Figure~\ref{fig:pumpprobe}) already indicates that magnetic dipole (M1) selection rules alone do not account for all the observed lines.
Selection rules between the hyperfine levels allow further investigation into the nature of this transition. Judd-Ofelt theory~\cite{Judd_1962, Ofelt_1962} allows the admixture of even parity states and \textit{forced dipole} (FD) transitions become allowed through these admixtures.
We model the FD transitions by admixture of the ground and excited states into $J = 5/2, 7/2, 9/2$ excited configurations coupled by an odd parity tensor operator $T_3^1$, which ensures the $\abs{\Delta m_j} = 3$ transition is strongly forbidden.
A multiparameter fit, including a Boltzmann factor, to the low-power fluorescence spectrum yield a best-fit combined M1 ($70\%$) and FD ($30\%$) transition with a fitted sample temperature of \SI{8.12}{\kelvin}. Relative amplitudes for M1 and FD transitions are shown in Figure~\ref{fig:hyperfine}.

Alternatively (or additionally), the dielectric medium surrounding the Tm atom may couple to the quadrupolar moments, allowing \textit{pseudoquadrupole} (PQ) transitions~\cite{Jorgensen_1964, Mason_1974}, which have been used to explain hypersensitive transitions of lanthanide ions in solution~\cite{GorllerWalrand_1998}.
A $Y_3^2$ (tetrahedral) distortion of the permittivity surrounding the defect couples the dipolar radiation field with the quadrupolar moments of the atom, while also enforcing zero amplitude for the outermost $\abs{\Delta m_F} = 0$ transitions which are strongly forbidden among the $\abs{m_J}: 1/2 \leftrightarrow 1/2$ hyperfine transitions (see Figure~\ref{fig:hyperfine}).
Fitting the fluorescence spectrum to M1 and PQ transition strengths and sample temperature yields a combined M1 ($86\%$) and PQ ($14\%$) transition at $T =\SI{6.99}{\kelvin}$.
However, the best fit value cannot reproduce the relative line strengths exactly, and there may be residual saturation or optical pumping effects that cause deviations in the observed relative strengths.
Supporting this, the relative line strengths determined by measuring visible fluorescence are not the same as those determined from IR fluorescence. Figure~\ref{fig:hyperfine} shows the expected relative amplitudes for pure M1, FD, and PQ transitions, alongside the observed hyperfine features.
We determine that a combination of M1 and PQ transitions best explain the observed selection rules.

Dielectric effects could also contribute to the reduced lifetime of the Tm excited ${}^2F_{5/2}$ state in the argon matrix, measured at \SI{30}{\milli\second} for Site I compared to \SI{112}{\milli\second} as measured in an optical lattice~\cite{Golovizin_2017}.
The discrepancy is partially explained by an $n^3$ enhancement of the emission rate due to photon flux and density of states corrections for M1 transitions~\cite{GorllerWalrand_1998, Rikken_1995}.
At \SI{4}{\kelvin}, the index of refraction of argon~\cite{Sinnock_1969} is $n = 1.29$, which should reduce the excited state lifetime via M1 to \SI{52}{\milli\second}, with further reductions possibly resulting from the inclusion of the FD or PQ transitions in the decay pathway.

\section{Zeeman Effect\label{sec:magnetic}}

\begin{figure}
    \centering
    \includegraphics[width=\linewidth]{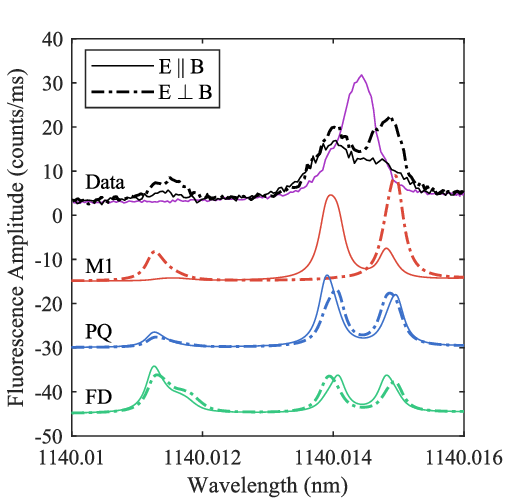}
    \caption{
        Polarized fluorescence spectrum obtained with a vertical magnetic field estimated at \SI{10}{\milli\tesla}.
        Underset are relative amplitudes of M1 (magnetic dipole), PQ (pseudoquadrupole), and FD (forced dipole) transitions, obtained with a Monte Carlo simulation for randomly oriented crystal axes.
        }
    \label{fig:polarized}
\end{figure}

The ${}^2 F_{7/2} \leftrightarrow {}^2F_{5/2}$ inner shell transition of thulium in solid argon features optical linewidths as low as \SI{0.56}{\pico\meter} (\SI{129}{\mega\hertz}), allowing direct optical detection of Zeeman splitting induced by an applied magnetic field.
Randomly oriented crystal axes in the disordered crystal experience linear or quadratic Zeeman splitting depending on their angle with respect to the applied field; with a field strength of \SI{10}{\milli\tesla}, this allows the spectral resolution of crystal axis orientations.
We use this effect to perform polarized absorption spectroscopy~\cite{Binnemans_2015, GorllerWalrand_1998} with the excitation polarized parallel or perpendicular to the applied field, and consequently oriented with respect to the crystal axis depending on the linear or quadratic Zeeman shift.
We study the magnetic and electric components of the transition using polarized spectroscopy for the subset of hyperfine lines in the $\abs{m_J} : 1/2 \leftrightarrow 1/2$ transition, shown in Figure~\ref{fig:polarized} for polarization parallel and perpendicular to the applied \SI{10}{\milli\tesla} field.
Expected line amplitudes for M1, FD, and PQ transitions are obtained for a simulation of $N=500$ defect sites with randomly oriented crystal axes and light polarization oriented with respect to the applied field to match the experiment, and are shown in Figure~\ref{fig:polarized}.
Observed line amplitudes from polarized excitation provide further evidence that the total transition includes FD or PQ components.

\begin{figure}
    \centering
    \includegraphics[width=\linewidth]{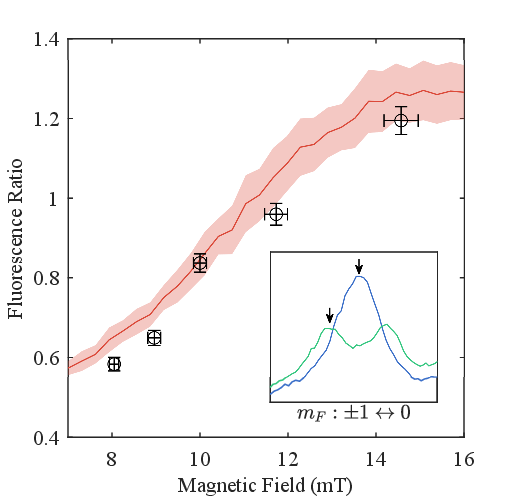}
    \caption{
        Ratiometric measurement between fluorescence amplitudes at \SI{1140.0141}{\nano\meter} and \SI{1140.0149}{\nano\meter} (shown in the inset) as a function of magnetic field strength, obtained for unpolarized light.
        The solid line and shaded region correspond to Monte Carlo simulations of randomly oriented crystal axes and linewidths (FWHM) of \SI{270}{\mega\hertz}.
        }
    \label{fig:sensitivity}
\end{figure}

We also investigate the system's use as a magnetometer using a two color measurement scheme at the center (\SI{1140.0149}{\nano\meter}) and inflection point (\SI{1140.0141}{\nano\meter}) of the $m_F: 0 \leftrightarrow \pm 1$ transition within the $|m_F| : 1/2 \leftrightarrow 1/2$ hyperfine group; the line splits into two components under a magnetic field.
At each measured field strength, the ratio of fluorescence decay amplitudes is calculated, providing a calibration-free estimate of the field, as shown in Figure~\ref{fig:sensitivity}.
Trends from Monte Carlo simulations with parameters chosen to closely match the data are also shown.
Simulations fail to match observed data exactly, likely due to imperfect choice of parameters (line widths, amplitudes, background) and ignorance of nearby background features.

In the range of \SI{10}{\milli\tesla} and near the shot-noise limit, we estimate the DC sensitivity to be \SI{330}{\micro\tesla\per\sqrt{\hertz}}, where the power equivalent bandwidth is determined from the total experiment time.
For reference, the dipole field from a single electron is \SI{\sim 1}{\micro\tesla} at a distance of \SI{10}{\nano\meter}.
While not currently competitive with the \SI{}{\nano\tesla\per\sqrt{\hertz}} sensitivities available using all-optical methods in NV centers~\cite{Wickenbrock_2016,Burgler_2023}, there are a number of paths which could increase the sensitivity by orders of magnitude.
In the shot-noise limit, the sensitivity is improved by a factor $\sqrt{N}$ of the fluorescence collected in a fixed time interval, which could be achieved with higher detection efficiency, cavity enhancement, or optical pumping out of the $J = 5/2$ level.
In addition, both the sensitivity and sensitive range are linearly dependent on the linewidth; even under current experimental limitations, a factor of $3$ improvement in the sensitivity is immediately available using the narrowest achievable lines and an improved detection duty cycle.
The ultimate limit of the linewidth is the transform limit of \SI{30}{\hertz}, which would offer an orders of magnitude improvement in the sensitivity.
Currently it is unknown how close to this limit can be reached: the inhomogeneous linewidth can be reduced with improved annealing strategies, and the homogeneous linewidth can be reduced by cooling.
We also note that fluorescence ratio metric is naturally less sensitive near zero field, however it is not uncommon to operate nanoscale magnetometers under a small bias field~\cite{Rondin_2014}.

\section{Conclusion}

We have used pump-probe spectroscopy, hyperfine structure, and selection rules to identify crystal field splitting of the inner-shell transition of thulium across two stable trapping sites in solid argon.
We also demonstrate the potential for measuring the magnitude of magnetic fields, using optically detected Zeeman shifts from \SI{}{\milli\tesla} fields with a DC sensitivity of \SI{330}{\micro\tesla\per\sqrt{\hertz}}.
While currently less sensitive than all-optical nanoscale magnetometers based on NV centers, this is the first detection of small fields in matrix-isolated atoms that avoids the need for microwave or radio frequency radiation.
In the future, use of selective bleaching methods or light-induced trapping site transfer, which has been observed in other matrix-isolated systems~\cite{Kanagin_2013, Tao_2015}, may allow the isolation of a single crystal axis and vector magnetometry.
The magnetic sensitivity is limited by the ensemble optical linewidth of \SI{0.56}{\pico\meter} (\SI{129}{\mega\hertz}). This was achieved with careful sample annealing, and is likely homogeneously broadened and limited by the same fast internal relaxation between crystal field levels that enables the pump-probe spectroscopy.
The homogeneous broadening could be reduced through additional cooling, and is expected to scale exponentially with inverse temperature in the lowest temperature regime.
Furthermore, the detailed knowledge of trapping sites revealed here allows more sophisticated studies of the annealing dynamics and possible improvement of inhomogeneous linewidths as well.

\begin{acknowledgments}
This work was supported by the National Science Foundation under Grant No. PHY-2310394.
\end{acknowledgments}

\section{Data Availability}
The data that supports the findings of this article are openly available~\cite{github}.

\end{document}